\documentclass[11pt]{iopart}
\usepackage{graphicx}
\usepackage{epsf,amsfonts,amssymb,amsbsy,wrapfig,cite}
\begin{document}
\title[Quantum black hole and Hawking radiation]
{Quantum black hole and Hawking radiation\\ at microscopic
magnifying}
\author{V.V.Kiselev}
\address{
Russian State Research Center ``Institute for
High Energy
Physics'', 
Pobeda 1, Protvino, Moscow Region, 142281, Russia}
\ead{kiselev@th1.ihep.su}
\begin{abstract}
We establish a state of stopping the Hawking radiation by quantum
Schwarzschild black hole in the framework of quasi-classical
thermal quantization for particles behind the horizon. The
mechanism of 
absorption and radiation by the black hole is presented.
\end{abstract}


\section{Introduction}

The Hawking radiation of black hole \cite{Hawking,Hawking2} has
got a deviation from thermal black-body spectrum of energy. A
reason for such the grey-body spectrum is the
quantum-mecha\-ni\-cal rescattering of particle propagating from
horizon to observer at infinity, by the gravitational potential of
black hole. Let us illustrate this point by the simplest case of
Schwarzschild black hole. The metric is defined by
\begin{equation}\label{intr1}
    {\rm d}s^2=g_{tt}(r)\,{\rm d}^2t-\frac{1}{g_{tt}(r)}\,{\rm
    d}r^2-r^2\,[{\rm d}\theta^2+\sin^2\theta\,{\rm d}\phi^2],
\end{equation}
with
\begin{equation}\label{intr2}
    g_{tt}(r)=1-\frac{r_g}{r},
\end{equation}
where $r_g$ is the radius of Schwarzschild sphere, i.e. the single
horizon of black hole.

For radial geodesics, the action on a trajectory can be
represented as
\begin{equation}\label{U-1}
    {S}_{HJ}=-{E}\,t+{\cal S}_{HJ}(r),
\end{equation}
where $E$ is a conserved energy of massive particle, that defines
the dimensionless integral of motion
\begin{equation}\label{U-2}
    A=\frac{m^2}{E^2}.
\end{equation}
The Hamilton--Jacobi equation reads off
\begin{equation}\label{U}
    \frac{1}{m^2}\,\left(\frac{\partial{\cal S}_{HJ}}{\partial
    r_*}\right)^2 = {\cal E}_A-U(r),\qquad {\cal E}_A =
    \frac{1}{A},\quad U(r)=g_{tt}(r)=1-\frac{r_g}{r},
\end{equation}
where the `tortoise' coordinate is defined by
\begin{equation}\label{U1}
    r_*=\int\frac{{\rm d}r}{g_{tt}(r)}
    =r+r_g\ln\left[\frac{r}{r_g}-1\right].
\end{equation}
The Hamilton--Jacobi equation of (\ref{U}) states the following:
the total energy ${\cal E}_A$ is the sum of kinetic term (radial
derivatives) and potential term $U$.

\begin{figure}[th]
  \centerline{\includegraphics[width=7cm]{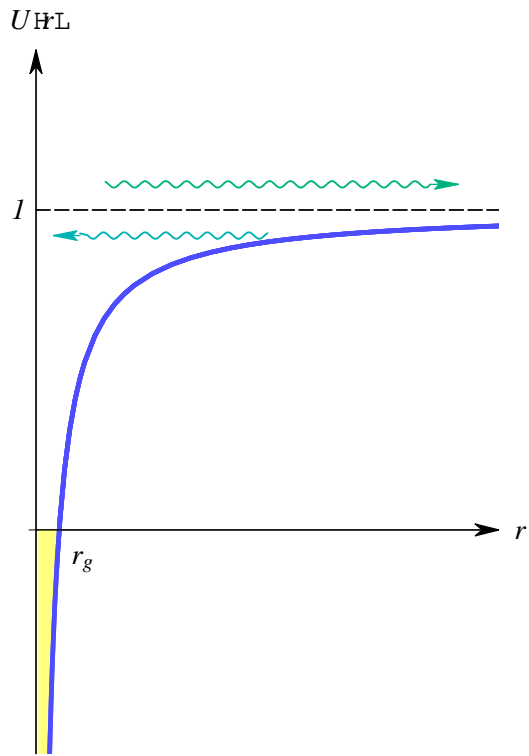}}
  \caption{The potential of radial motion and quantum reflection of particle
  travelling to infinity.}\label{Hpic1}
\end{figure}

The problem of radial motion is illustrated by Fig. \ref{Hpic1},
wherein we have depicted the `potential' and reflection of
outgoing waves, which are shown schematically\footnote{The
reflected wave has the same energy as the outgoing wave, of
course.}.

\noindent
 Therefore, there is a nonzero coefficient of reflection, which
 causes the grey-body factors depending on the particle energy.
 The Hawking radiation originates from
 vacuum fluctuations in vicinity of horizon. The virtual fluctuations
 are transformed to observed particles due to action of
 gravitational field. The pair of particles in fluctuation is
 separated by the field in the following way: the positive energy particle goes
 to infinity, while the negative energy particle falls behind the
 horizon diminishing the total mass of black hole in the exact
 balance with the radiated energy.

In actual analogy with reasons for the appearance of grey-body
factors, let us make the following question: Can the rescattering
of falling particle lead to its reflection by the black hole? If
such the reflection does exist, it could suppress the absorption
of particle by the black hole, and, hence, particularly the
Hawking radiation, too. Moreover, the total reflection would
result in the complete stopping of Hawking radiation.

In order to answer this question we use the quasi-classical
thermal approach to the space-time behind the horizon as it was
recently developed in \cite{K1,K2}. The introduction of such new
framework is required by the consideration of trajectories
completely confined behind the horizon. Eq. (\ref{U}) and Fig.
\ref{Hpic1} clearly show that mentioned trajectories
cor\-re\-spond to
\begin{equation*}
    {\cal E}_A \leqslant 0,
\end{equation*}
implying imaginary values of energy and time treated as the
indication of statistical description. Namely, the period of
motion in imaginary time equals the inverse tem\-pe\-rature of
particles in thermal ensemble existing behind the horizon. The
thermal equilibrium demanding the exact periodicity leads to the
quasi-classical quantization \mbox{\textit{a la}} the introduction
of old Bohr orbits for the particles confined behind the horizon.
In the present paper we investigate the ground quantum state of
Schwarzschild black hole and its excitations in detail. We show
how transitions between the excitations lead to the Hawking
radiation with the Gibbs distribution in energy
\begin{equation*}
    w\sim {\rm e}^{-\beta\,E},
\end{equation*}
where $\beta=1/T$ is the inverse temperature of black hole, and
$E$ is the energy of radiation quantum. Then we find the
condition, when the Hawking radiation is stopped: the black hole
finishes to radiate, if all the excitations have transited to the
ground state.

The paper is organized as follows: In section 2 we remind basic
points in classifying the radial trajectories confined behind the
horizon and develop the analysis by separating regular
contributions described in \cite{K1,K2} and so-called ``mute''
terms, which are analogous to the Dirac sea, since they
are cancelled in the total energy and partition function. Then we
apply the energy conservation in order to describe excitations.
The excitations are formed by energetic particles coupled with
their ``antipodes'', which have an opposite sign of Euclidean
energy. In section 3 we present the mechanism of particle
absorption by the black hole and the limit of large quantum
numbers for the Hawking radiation, which, by derivation, obeys the
Gibbs distribution. The ground state of quantum black hole does
not produce any Hawking radiation. Section 4 summarizes our
conclusions.

\section{Quantum levels}
The causal geodesics of massive particle confined behind the
horizon with time-like intervals can be described in terms of
metric for the Schwarzschild black hole \cite{K1}
\begin{equation}\label{euclid-s}
    {\rm d}s^2=\frac{r_g}{r}\cdot e^{\displaystyle-\frac{r}{r_g}}\cdot({\rm
    d}\rho^2+\rho^2{\rm d}\varphi_E^2)-r^2({\rm d}\theta^2+\sin^2\theta{\rm
    d}\phi^2),
\end{equation}
where
\begin{equation}\label{euclid-def2}
    \left\{\begin{array}{rlrl}
t = & -{\rm
 i}\,2r_g\,\varphi_E,&\qquad \varphi_E\in & [0,2\pi],
 \\[4mm]\displaystyle
 r_*= & \displaystyle2 r_g \ln\left[-\frac{\rho}{2 r_g}\right],
  & \qquad\rho \in & [0,2r_g],\end{array}
 \right.
\end{equation}
and the Euclidean phase $\varphi_E$ has a period equal to $2\pi$,
corresponding to the inverse temperature
\begin{equation}\label{beta}
    \beta =4\pi r_g.
\end{equation}
The trajectory is determined by
\begin{equation}\label{derE}
    \frac{{\rm d}\varphi_E}{{\rm d}r} =
    \frac{1}{2r_g}\,\sqrt{1-\frac{r_c}{r_g}}\,\frac{r}{r_g-r}\,\sqrt{\frac{r}{r_c-r}},
\end{equation}
where the maximal remoteness of particle from the singularity at
$r=0$ is
\begin{equation}\label{rc}
    r_c=-r_g\,\frac{A}{1-A}\leqslant r_g,\qquad \mbox{at } A<0.
\end{equation}

A \textit{regular} cycle is determined by (\ref{derE}), and its
value for $\varphi_E$ is equal to the phase increment during
motion from the singularity to $r_c$ and back
\begin{equation}\label{cycle}
    \Delta_c\varphi_E = 2\int\limits_{0}^{r_c}{\rm d}r\,\frac{{\rm d}\varphi_E}{{\rm
    d}r}=
    \frac{\pi}{2}\,
    \bigg[2-(2+x)\sqrt{1-x}\bigg],\qquad x=\frac{r_c}{r_g}.
\end{equation}
A new feature we involve is a so-called ``mute'' solution of
(\ref{derE}) defined by
\begin{equation}\label{mute1}
    r\equiv r_c, \qquad\mbox{at } r_c < r_g,
\end{equation}
which is possible because of singularity in (\ref{derE}), since
\begin{equation}\label{mute2}
    \frac{{\rm d}r}{{\rm d}\varphi_E}\equiv 0, \qquad\mbox{at }
    r\equiv r_c< r_g.
\end{equation}

\begin{figure}[th]
  \centerline{\includegraphics[width=6cm]{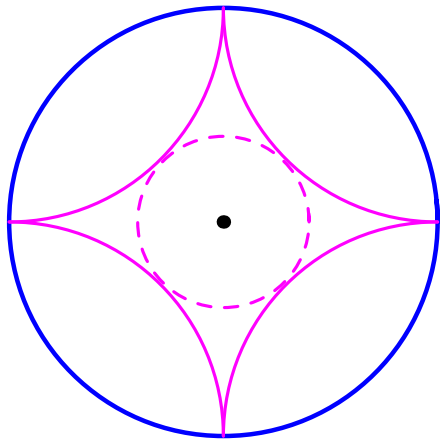}}
  \caption{The geodesics confined behind the horizon in
    polar coordinates $\{\rho,\varphi_E\}$: the winding number
    equal to 4 is shown by solid arcs, the ``mute'' trajectory with the same $r_c$ is
  depicted by the dashed circle. The central point corresponds to the horizon
  at $r=r_g$, while the solid circle is the singularity.}\label{Hpic2}
\end{figure}

If we ignore the ``mute'' solution, then the regular number of
cycles per period is naively given by
\begin{equation}\label{reg-n}
    \tilde n =\frac{2\pi}{\Delta _c\varphi_E},
\end{equation}
which we call the ``winding number''. The examples of geodesics
are shown in Fig. \ref{Hpic2}.

\begin{figure}[th]
  \centerline{\includegraphics[width=6cm]{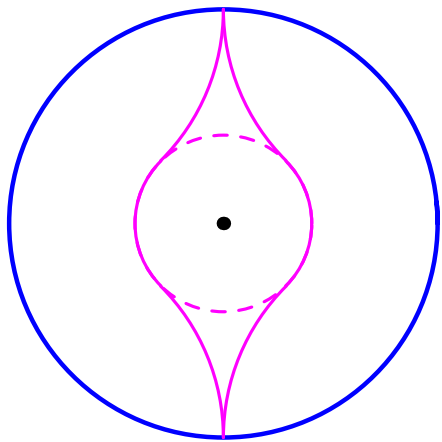}}
  \caption{The mixed geodesics containing ``mute'' parts.}\label{Hpic3}
\end{figure}

However, due to the ``mute'' geodesics the particle can get a
mixed regular-mute trajectories as shown in Fig. \ref{Hpic3},
since both geodesics are tangent at maximal distance from the
singularity. We can say that the regular trajectory can adhere to
the ``mute'' path at $r=r_c$. Therefore, one should determine a
role of ``mute'' geodesics.

In this way, let us, first, calculate the action on the mute
trajectory (${\rm d}r\equiv0$):
\begin{equation}\label{mute3}
    S_{\rm mute} = - {\cal E}\oint{\rm d}t_E,
\end{equation}
where the `Euclidean' time is given by
\begin{equation}\label{tE}
    t_E=\frac{\beta}{2\pi}\,\varphi_E,
\end{equation}
while ${\cal E}$ is the `Euclidean' energy, so that $A=-m^2/{\cal
E}^2$. Then, the action per period is
\begin{equation}\label{mute4}
    S_{\rm mute} = - \beta\,{\cal E}.
\end{equation}
Emphasize, the ``mute'' action depends on the sign of energy $\cal
E$.

Second, let us exploit the energy conservation. Then, the total
energy of black hole is its mass $M$, which is a real number. The
energy $\cal E$ represents the imaginary contribution to the total
energy. Therefore, this contribution should be cancelled by
opposite term. So, we \textbf{postulate} that for each particle on
the ``mute'' geodesics there is the ``anti-mute'' particle with
opposite energy. Thus, the sum of actions will nullify:
\begin{equation}\label{mute5}
    S_{\rm mute}+S_{\mbox{\scriptsize anti-mute}} \equiv 0.
\end{equation}
Therefore, the ``mute''---``anti-mute'' pair does not contribute
to the total energy as well as to the partition function: the mute
terms are cancelled. Then, we can omit the mute part from the
action in a full analogy with the Dirac sea.

However, ``mute'' pathes can contribute indirectly. Indeed, for
the regular term, the increment of action per cycle is given by
\begin{equation}\label{L1}
    \Delta_c S = - m\,\pi\,r_g\,x^{3/2}.
\end{equation}
Note, it is independent on the sign of energy $\cal E$. Then, the
action per period is determined by the increment and the number of
cycles corrected by the phase belonging to the mute part,
\begin{equation}\label{L2}
    n =\frac{2\pi-\Delta_{\rm mute}\varphi_E}{\Delta_c\varphi_E},
\end{equation}
so that
\begin{equation}\label{L3}
    S=-m\,\pi\,r_g\,x_n^{3/2}\,n,
\end{equation}
where we explicitly mark the maximal remoteness by the natural
index of actual winding number in terms of $x$: $r_c= x_n\,r_g$.
The corresponding term of mute path at the trajectory is given by
\begin{equation}\label{L4}
    \Delta S_{\rm mute}=-\beta\,{\cal E}\,\frac{\Delta_{\rm
    mute}\varphi_E}{2\pi}.
\end{equation}
By energy conservation, we \textbf{postulate} that the same
trajectory with the opposite sign of energy should accompany the
particle. For such the mirror trajectory, we call the
``antipode''. Then, the ``mute'' terms are cancelled again, while
the regular contribution of particle--antipode pair is double:
\begin{equation}\label{L5}
    S_{\rm pair} =-2m\,\pi\,r_g\,x_n^{3/2}\,n.
\end{equation}
This postulate has two items.
\begin{description}
    \item[\textit{i)~}] At the ground level, ${\cal E}\equiv 0$,
    particles and their antipodes are indistinguishable; there are
    no mute pathes.
    \item[\textit{ii)}] Excitations of ground level exist in the
    form of ``particle--antipode'' pairs.
\end{description}

For definiteness, we have to fix the value of phase shift
$\Delta_{\rm mute}\varphi_E$. A spectacular way is making use of
correspondence principle: at large winding numbers the
quasi-classical description has to match with the classical
dynamics. In our case this principle implies the following:

The limit of $n\gg 1$ corresponds to $x\to 0$. Then, the phase
increment is given by
\begin{equation}\label{L6}
    \Delta_c\varphi_E \to \frac{3\pi}{8}\,\frac{1}{x^2},
\end{equation}
while
\begin{equation}\label{L7}
    m =|{\cal E}|\sqrt{-A} \to |{\cal E}|\,x^{1/2}.
\end{equation}
Therefore,
\begin{equation}\label{L8}
    S_{\rm pair} = - \beta\,{\cal E}_{\rm
    pair}\cdot\frac{4}{3}\,\frac{2\pi-\Delta_{\rm
    mute}\varphi_E}{2\pi},
\end{equation}
where the positive energy of particle--antipode pair is defined by
\begin{equation}\label{L9}
    {\cal E}_{\rm pair} =2\,|{\cal E}|.
\end{equation}
As we have demonstrated in \cite{K1,K2}, the action at such the
trajectories with imaginary time contributes to the partition
function in the standard way:
\begin{equation}\label{L10}
    w={\rm e}^{S},
\end{equation}
which should restore the Gibbs distribution in the classical limit
of $n\gg 1$,
\begin{equation*}
    w={\rm e}^{-\beta\,E}.
\end{equation*}
Then, the correspondence principle dictates
\begin{equation}\label{L11}
    \Delta_{\rm mute}\varphi_E = \frac{\pi}{2},\qquad \mbox{at }
    r_c<r_g,
\end{equation}
where we remind that this prescription is valid for the excited
levels, while at the ground level $\Delta_{\rm mute}\varphi_E
\equiv 0$, since there are no ``mute'' trajectories for massive
particles at $r=r_g$ (the interval is exactly equal to zero).

At the moment, we can easily consider two limits in calculating
the partition function.
\begin{enumerate}
    \item[1.] If all particles are at the ground level, then the sum
    of actions is reduced to the sum of particle masses
    \begin{equation*}
    G = \sum S= -\frac{\beta}{2}\sum m =-\frac{1}{2}\,\beta\,M,
\end{equation*}
where $M$ is the black hole mass:
\begin{equation*}
    M=\sum m.
\end{equation*}
The thermodynamical function $G$ reproduces the correct value for
the product of inverse temperature to the free energy ${\cal F}$
\begin{equation*}
    G = -\beta\,{\cal F},\qquad {\cal F}=\frac{1}{2}\,M.
\end{equation*}
    \item[2.] If all particles are highly excited, then the sum of
    energies for the particle--antipode pairs gives the black hole
    mass again;
    \begin{equation*}
    M=\int\limits_{0}^{M}\,{\rm d}{\cal E}_{\rm pair},
\end{equation*}
where we have introduced the notation of differential for the
energy of single pair, which is infinitely small in comparison
with the total energy: ${\cal E}_{\rm pair}\mapsto {\rm d}{\cal
E}_{\rm pair}$, and we insert the integration instead of sum over
pairs. In that case, the temperature is determined by the summed
energy of other pairs, so that
\begin{equation*}
    \beta =8\pi\,M\mapsto \beta({\cal E}_{\rm pair})=8\pi\,{\cal E}_{\rm
    pair},
\end{equation*}
hence, the $G$ function is obtained by the integration
\begin{equation*}
    G=-\int\limits_0^M\,\beta({\cal E}_{\rm pair})\,{\rm d}{\cal E}_{\rm
    pair}=-4\pi\,{\cal E}^2_{\rm
    pair}\Big|^M_0=-\frac{1}{2}\,\beta\,M,
\end{equation*}
which again reproduces the correct value.
\end{enumerate}
One can easily recognize that the mixed situation can be described
as a simple combination of two limits above. It is spectacular,
that the mass of black hole is given by the sum of both masses for
particles at the ground level and energies of particles and
antipodes at excited levels.

\begin{figure}[th]
  \setlength{\unitlength}{1mm}
  \begin{center}
  \begin{picture}(90,60)
  \put(0,0){\includegraphics[width=9cm]{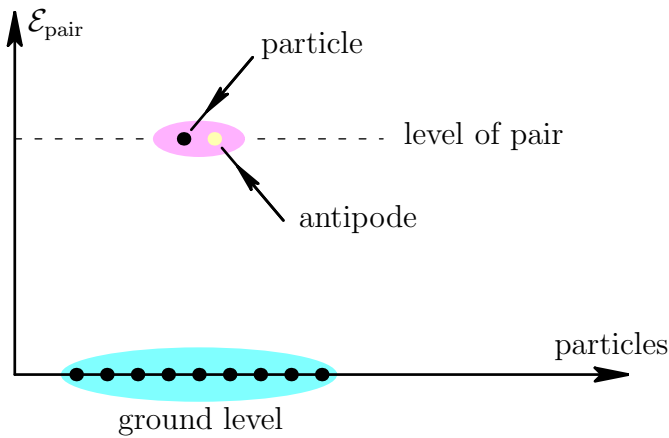}}
  \put(22,0){\mbox{ground level}}
  \put(80,10){\mbox{particles}}
  \put(10,53){${\cal E}_{\rm pair}$}
  \put(41,50){\mbox{particle}}
  \put(46,27){\mbox{antipode}}
  \put(60,38){\mbox{level of pair}}
  \end{picture}
  \end{center}
  \caption{Quantum structure of Schwarzschild black hole.}\label{Hpic4}
\end{figure}

We show the schematic structure of quantum black hole in Fig.
\ref{Hpic4}.

Finally, let us present a simple analogy for the explanation of
phase shift at excited levels and its absence at the ground level.
We start from the statement on the periodic motion
\begin{equation}\label{P1}
    \oint {\rm d}\varphi_E =2\pi.
\end{equation}

\noindent
 In order to show the analogy with the quasi-classical
 quantization we can multiply by the winding number, so that
\begin{equation}\label{P2}
    \oint\,n\cdot{\rm d}\varphi_E=2\pi\,n.
\end{equation}
This relation supports the interpretation of $\varphi_E$ as the
dynamical variable and $n$ as the canonically conjugated momentum
of $\varphi_E$.

Next, we isolate the regular term, which is given by the
integration over the distance,
\begin{equation}\label{P3}
    \oint\limits_{\rm reg.}n\cdot{\rm d}\varphi_E=n\cdot
    2n\int\limits_0^{r_c}\frac{{\rm d}\varphi_E}{{\rm d}r}\,{\rm
    d}r=n^2\cdot\Delta_c\varphi_E.
\end{equation}
This regular term is a complete analogue of complex phase for the
wavefunction in the quasi-classical approximation,
\begin{equation*}
    \oint\limits_{\rm reg.}{\cal P}{\rm d}{\cal Q},
\end{equation*}
where $\cal Q$ is a dynamical variable, and $\cal P$ is its
momentum.

Then, Fig. \ref{Hpic1} shows that each reflection of trajectory at
$r=r_c$ gives the additional phase equal to $-\pi/2$ for the
reflected wave with respect to the complex phase of wave falling
to $r=r_c$, and the additional phase equal to $\pi$ for the
reflection at $r=0$, since the potential wall at $r\leqslant 0$ is
infinitely high. The quasi-classical quantization by
Bohr--Sommerfeld results in
\begin{equation}\label{P4}
    \oint\limits_{\rm reg.}n\cdot{\rm
    d}\varphi_E=2\pi\,n+n\cdot\left(\frac{\pi}{2}-\pi\right)=n\cdot\frac{3\pi}{2}.
\end{equation}
Therefore, the true winding number at $r_c<r_g$ is given by
\begin{equation}\label{P5}
    n =\frac{3\pi}{2\Delta_c\varphi_E}.
\end{equation}
The quantization by Bohr--Sommerfeld is quite accurate, if $r_c\ll
r_g$ because of the following reason: In fact, we expect that the
wavefunction should nullify at $r=r_g$, which is possible, if both
rising and falling exponents contribute at $r>r_c$. However, we
can neglect the rising exponent, which compensates the tail of
falling exponent at $r=r_g$, if $r_c\ll r_g$. Then, the rules
applied are justified.

The situation is slightly changed for the ground state, since the
wavefunction should nullify at $r=r_g$, which implies the complete
reflection of wave at right return point. The corresponding phase
between falling and reflected waves is equal to $-\pi$ instead of
$-\pi/2$. Then, the regular equation for the winding number at the
ground state is restored,
\begin{equation}\label{P6}
    n_{\rm ground} =\frac{2\pi}{\Delta_c\varphi_E},
\end{equation}
in a full agreement with the consideration in \cite{K1,K2}.

The above study supports the following representation of
wavefunction at $r<r_c<r_g$:
\begin{equation}\label{P7}
    \Psi_{r>0}\sim +\cos\left(n\int\limits_0^r \frac{{\rm d}\varphi_E}{{\rm d}r}\,{\rm
    d}r+\frac{\pi}{2}\right),
\end{equation}
\begin{equation}\label{P7'}
    \Psi_{r<r_c}\sim -\cos\left(n\int\limits_{r_c}^r \frac{{\rm d}\varphi_E}{{\rm d}r}\,{\rm
    d}r+\frac{\pi}{4}\right),
\end{equation}
which are identical at
\begin{equation}\label{P8}
    \frac{1}{2}\,n\Delta_c\varphi_E=\pi+\left(\frac{\pi}{4}-\frac{\pi}{2}\right),
\end{equation}
that restores the quantization rule. The modification for the
ground state is transparent.

This consideration provides us with the evidence for the
reasonable treatment of quantum levels in the black hole.

\begin{figure}[th]
  \setlength{\unitlength}{1mm}
  \begin{center}
  \begin{picture}(110,70)
  \put(10,0){\includegraphics[width=10cm]{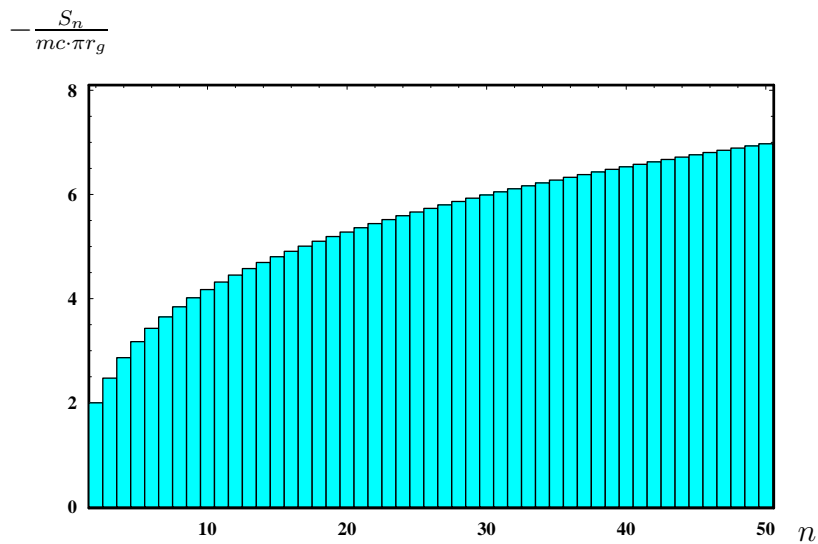}}
  \put(5,67){$-\frac{S_n}{mc\cdot \pi r_g}$}
  \put(110,0){$n$}
  \end{picture}
  \end{center}
  \caption{The quantum action.}\label{pic4-corr}
\end{figure}

In accordance with (\ref{L5}), (\ref{P5}) and (\ref{P6}), the
quantum action per single particle takes discrete values shown in
Fig. \ref{pic4-corr}, that slightly corrects naive Fig. 5 of
\cite{K1} because of modified winding number.

Thus, the quantum black hole is composed by the particles
occupying the ground levels as well as the excited levels. The
number of excitations depends on initial conditions before the
collapse, which determine the mass of black hole. The spectrum of
levels is discrete, though at high energy of excitation one could
apply the classical description with the Gibbs distribution.

\section{Absorbtion and Hawking radiation}

Let us consider a change of state for the quantum black hole due
to absorption of external particle falling behind the horizon.
Such the particle has a positive total energy (with respect to
observer at $r=\infty$). The basic point is the conservation of
energy: the energy of falling particle increases the mass of black
hole. We suggest that falling to singularity causes the change of
black hole mass and a transition of the particle to a quantum
level. The transition depends on the energy of particle.

\begin{figure}[th]
  \setlength{\unitlength}{1mm}
  \begin{center}
  \begin{picture}(130,90)
  \put(0,0){\includegraphics[width=13cm]{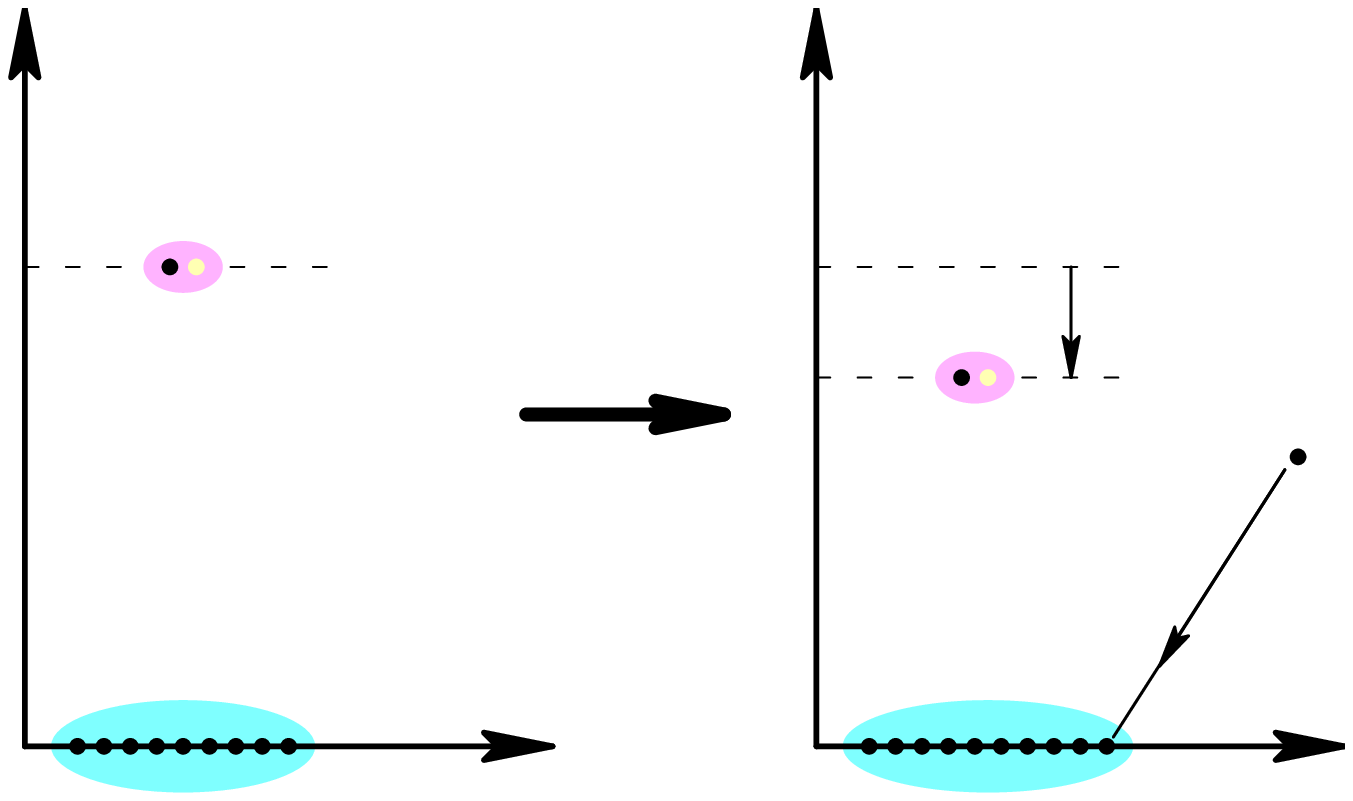}}
  \put(10,2){\mbox{ground level}}
  \put(45,14){\mbox{particles}}
  \put(10,78){${\cal E}_{\rm pair}$}
  \put(121,40){\mbox{particle}}
  \put(122,44){\mbox{falling}}
  \put(37,55){\mbox{level of pair}}
  \end{picture}
  \end{center}
  \caption{The mechanism of particle absorption at $E<m$.
  }\label{Hpic5}
\end{figure}

First, consider the case of $0<E<m$, i.e. the particle, which
binding energy is so large, that the maximal distance from the
horizon is finite. Then, the mechanism of absorption is shown in
Fig. \ref{Hpic5}: the falling particle occupies the ground level,
while the excited particle--antipode pair gets a transition to the
lower level.

\begin{figure}[th]
  \setlength{\unitlength}{1mm}
  \begin{center}
  \begin{picture}(130,90)
  \put(0,0){\includegraphics[width=13cm]{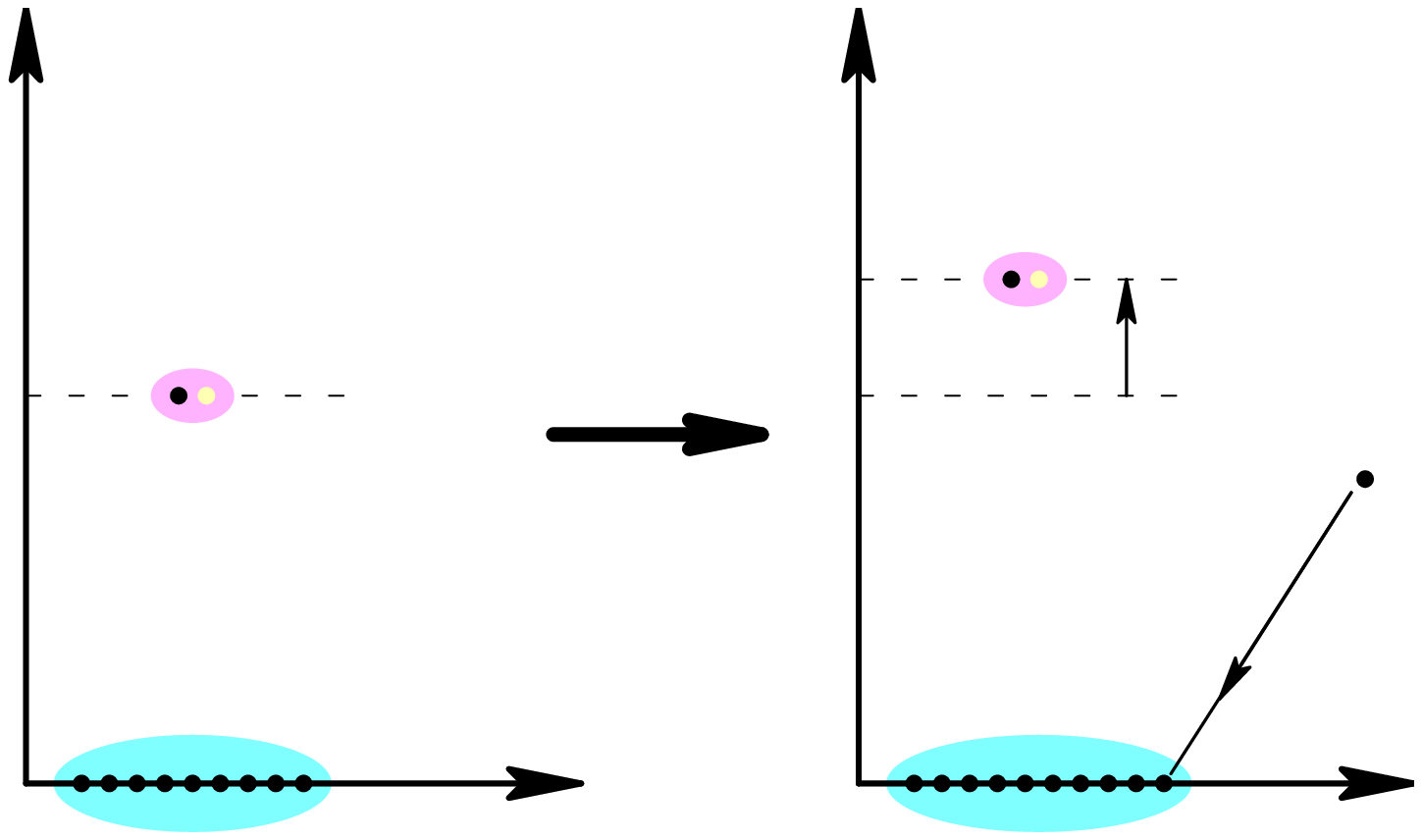}}
  \put(10,2){\mbox{ground level}}
  \put(45,14){\mbox{particles}}
  \put(10,78){${\cal E}_{\rm pair}$}
  \put(121,40){\mbox{particle}}
  \put(122,44){\mbox{falling}}
  \end{picture}
  \end{center}
  \caption{The mechanism of particle absorption at $E>m$.
  }\label{Hpic6}
\end{figure}

\noindent
 Due to energy conservation, the shift of level is given by the
 difference of black hole mass change and the mass of absorbed
 particle:
 \begin{equation}\label{Ab1}
    {\cal E}-{\cal E}'={\rm d}M-m=-{\rm d}{\cal E}, \qquad {\rm
    d}M=E.
\end{equation}
The partition function gets the correct change, of course: the
entropy increases by the appropriate value.

\begin{figure}[th]
  \setlength{\unitlength}{1mm}
  \begin{center}
  \begin{picture}(130,90)
  \put(0,0){\includegraphics[width=13cm]{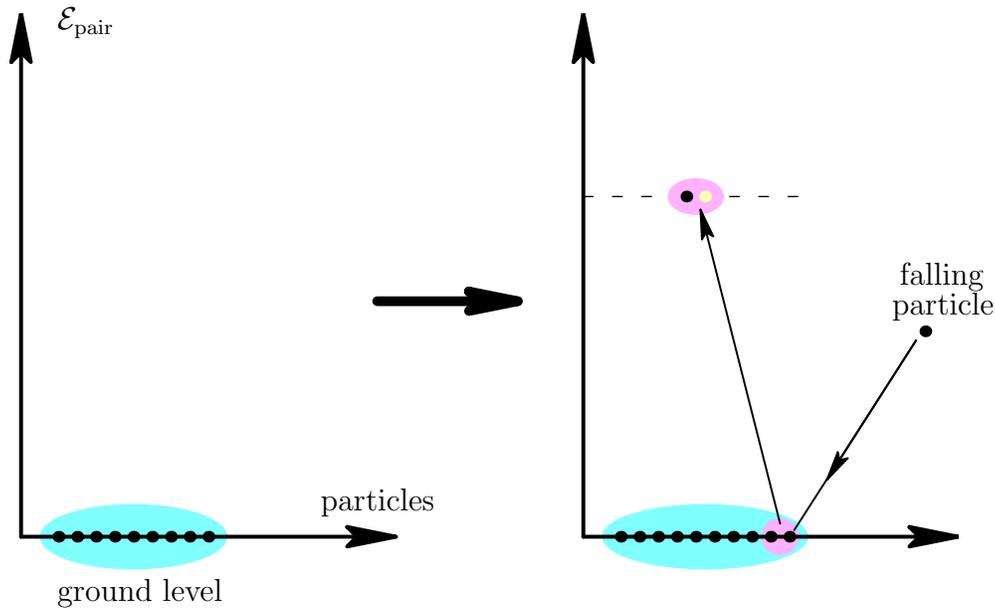}}
  \put(10,2){\mbox{ground level}}
  \put(45,14){\mbox{particles}}
  \put(10,78){${\cal E}_{\rm pair}$}
  \put(121,40){\mbox{particle}}
  \put(122,44){\mbox{falling}}
  \end{picture}
  \end{center}
  \caption{The mechanism of particle absorption at $E\gg m$.
  }\label{Hpic7}
\end{figure}

The case of $E>m$, i.e. the particle, which can have a nonzero
velocity at infinity, is quite analogous: the excited level gets a
transition to a higher one (see Fig. \ref{Hpic6}). In the limit of
extra large energy $E\gg m$, one could neglect the discreteness of
levels, so that the falling particle could excite an antipode from
the ground level in order to form a pair at large energy as shown
in Fig. \ref{Hpic7}.

Quantum restrictions to the absorption are quite evident. First,
we have the discrete spectrum of absorption, and this fact is
especially important at $E\sim m$. Second, the quantum black hole
will not absorb a particle with $E<m$, \textit{if the black hole
is completely at the ground level}, i.e. if there is no
excitations. Then, the quantum black hole can totally reflect the
falling particle, if the black hole is at the ground level. Even
at $E>m$, the energy could be too low in order to excite a higher
level of particle--antipode pair from the ground state. As we have
already mentioned in Introduction, such the total reflection
should affect the Hawking radiation, too.

Indeed, at $E> m$ we can simply invert the arrows in Figs.
\ref{Hpic6} and \ref{Hpic7} in order to get the process of
radiation\footnote{Inverting the process in Fig. \ref{Hpic5} could
be forbidden, if the particle--antipode pair is excited from the
ground state because of the energy balance: loosing the particle
of mass $m$ from the ground level should exceed the increase of
excitation energy, which could be not possible for some excited
levels. However, Fig. \ref{Hpic5} presents an alternative
mechanism of particle emission: if the quantum black hole has many
excitations, particles at the ground level could evaporate by
exciting the existing higher levels. However, the emitted
particles would fall back to the black hole, since their energies
are restricted by the mass, $E<m$.}. Then, the falling particle is
inverted to the outgoing particle of Hawking radiation reaching an
observer at infinity.

At $E\gg m$, we can easily estimate the probability of radiation,
since it is equal to the probability that the particle--antipode
pair \textit{was} at the excited level, which is equal to
\begin{equation}\label{A2}
    w={\rm e}^{-\beta\,{\cal E}_{\rm pair}},
\end{equation}
while the energy of pair is transmitted to the radiated quantum
(we neglect the mass of the particle):
\begin{equation}\label{A3}
    E\approx{\cal E}_{\rm pair},\qquad {\rm d}M\approx-{\cal E}_{\rm pair}.
\end{equation}
Therefore, we find the Gibbs distribution for the radiation of
single quantum, that leads to the black-body spectrum of
radiation, corrected by appropriate grey-body factors because of
rescattering on the gravitational potential\footnote{The Hawking
radiation as a tunnel-effect of pair creation by the gravitational
field in the quasi-classical approximation was considered in
\cite{Wil}.}.

The consequence of such quantum mechanism for the Hawking
radiation is clear: the radiation is completely stopped, if the
quantum black hole is at the ground level!

Highly excited levels correspond to classical description of black
hole, since particles move in the very vicinity to the
singularity. The ground state is extremely coherent: particles
homogeneously occupy all of the space behind the Schwarzschild
sphere. Such the coherence could result in the holographic state
\cite{Holograph}: the  state on the horizon sphere is equivalent
to the quantum state of whole black hole. However, the strict
statement requires an exact quantum theory of black holes, not the
quasi-classical approximation, although we expect that the
approximation used presents a rather valid qualitative picture for
the quantum black hole.

\section{Conclusion}

In short, we have described the inner quantum structure of
Schwarzschild black hole in the framework of quasi-classical
thermal approach. A particle in the thermal ensemble has the
ground state as well as excitations formed by a particle--antipode
pair. The antipode has the opposite sign of energy with respect to
the particle, that follows from the energy conservation. The
existence of ground state leads to stopping of Hawking radiation,
after all excited states have decayed to the ground level. We have
studied the mechanism of particle absorption and emission. The
Gibbs distribution for the excitations has been obtained.

We have considered the absorption and radiation of massive
particles by hot black hole. What modifications have to be
introduced, if we would involve neutral massless particles? We
expect that transitions of excited levels for massive particles to
lower ones will produce the Hawking radiation of such neutral
massless particles, until all excitations decay to the ground
level, again. Then, the radiation will stop, since no energy can
be extracted from the ground quantum state of black hole. In this
way, we get two rather dim items: a mechanism of energy transition
from massive particles to massless ones, and a sense of quantum
orbits for the massless particles inside the black hole.

This work is partially supported by the grant of the president of
Russian Federation for scientific schools NSc-1303.2003.2, and the
Russian Foundation for Basic Research, grant 04-02-17530.


\end{document}